\documentclass[trackchanges,twocolumn]{aastex701}

\begin{document}

\title{Constraining the Galactic bar and spiral pattern speeds with the Hyades tidal stream}

\correspondingauthor{Long Wang}
\email{wanglong8@sysu.edu.cn}

\author[orcid=0009-0009-7509-3794]{Zi-yi Zhou}
\affiliation{School of Physics and Astronomy, Sun Yat-sen University, Daxue Road, 519082 Zhuhai, China}
\email{zhouzy83@mail2.sysu.edu.cn}

\author[orcid=0000-0001-8713-0366]{Long Wang}
\affiliation{School of Physics and Astronomy, Sun Yat-sen University, Daxue Road, 519082 Zhuhai, China}
\affiliation{CSST Science Center for the Guangdong-Hong Kong-Macau Greater Bay Area, 519082 Zhuhai, China}
\email{wanglong8@sysu.edu.cn}

\author{Tereza Jerabkova}
\affiliation{Department of Theoretical Physics and Astrophysics, Faculty of Science, Masaryk University, Kotl\'{a}\v{r}sk\'{a} 2, Brno 611 37, Czech Republic}  
\email{tereza.jerabkova@sci.muni.cz}

\author{Zhenghao He}
\affiliation{School of Physics and Astronomy, Sun Yat-sen University, Daxue Road, 519082 Zhuhai, China}
\email{hezhh59@mail2.sysu.edu.cn}

\begin{abstract}

We present a suite of direct $N$-body simulations of the Hyades open cluster and its tidal stream in a Milky Way potential that includes a rotating bar and spiral arms. Using the high-resolution code \texttt{PETAR} and an \texttt{AGAMA}-based multi-component Galactic model, we vary the bar and spiral pattern speeds ($\Omega_b$, $\Omega_s$) on a discrete grid and quantify the resulting changes in stream orientation, length, and internal density structure. We compare the simulations to \textit{Gaia} EDR3 using the convergent point (CP) and compact convergent point (CCP) methods, followed by an adaptive three-dimensional nearest-neighbor matching in Cartesian space $(x,y,z)$. The \textit{Gaia} candidate members exhibit a pronounced longitudinal density peak at $Y_{\mathrm{rot}} \approx 0.1\,\mathrm{kpc}$ in a stream-aligned coordinate system. Models with $\Omega_s = 22.5\,\mathrm{km\,s^{-1}\,kpc^{-1}}$ and $\Omega_b \simeq 40$--$45\,\mathrm{km\,s^{-1}\,kpc^{-1}}$ best reproduce this feature, while faster-bar models fail to match the observed density structure. These models are consistent with recent constraints favoring a relatively slow Galactic bar, and they illustrate how nearby open-cluster streams can provide an independent, local constraint on non-axisymmetric Galactic dynamics.

\end{abstract}



\section{Introduction} 

Constraining the Milky Way gravitational potential is central to connecting stellar kinematics to the Galaxy's mass distribution and secular evolution \citep{MW_review..2016ARA&A..54..529B}. Even in the solar neighborhood, degeneracies persist between the halo shape and the disk and bulge/bar contributions. While the dark matter halo is often approximated by a spherical NFW profile \citep{NFW..1997ApJ...490..493N}, its precise geometry remains a subject of ongoing debate. Many recent analyses favor a flattened, oblate configuration ($q < 1$) to better match kinematic constraints \citep{Huang..2024NatAs...8.1294H}, although alternative models suggesting a prolate shape have also been proposed \citep{prolate_halo..2026A&A...706A.193Z}. In the disk, additional uncertainty arises from non-axisymmetric structure: the Galactic bar pattern speed remains debated, with ``fast'' ($\sim 55\,\mathrm{km\,s^{-1}\,kpc^{-1}}$) and ``slow'' ($\sim 40\,\mathrm{km\,s^{-1}\,kpc^{-1}}$) solutions supported by different dynamical tracers \citep{fast_bar..1999ApJ...524L..35D, slow_bar..2019MNRAS.488.4552S, slow_bar..2019MNRAS.489.3519C, bar_speed..2024MNRAS.531L..14L, bar_speed_review..2025NewAR.10001721H}. The morphology and dynamical nature of the spiral arms are likewise uncertain \citep{spiral_arms..2012MNRAS.421.1529G, Hunter..2024A&A...692A.216H, non_asy_potential..2025A&A...699A.263K}.

Stellar tidal streams---formed by the gradual stripping of star clusters---provide phase-coherent tracers of the Galactic potential \citep{stream_estimate_potential..2010ApJ...712..260K, Gibbons..2014MNRAS.445.3788G}. Globular-cluster streams are typically narrow and prominent \citep[e.g.,][]{Pal5..2001ApJ...548L.165O}, but they often probe the potential at larger Galactocentric distances. In contrast, open-cluster streams reside in the disk and are therefore particularly sensitive to perturbations from the bar, spiral arms, and encounters with giant molecular clouds \citep{stream_with_collision..2006MNRAS.371..793G, OLR..2016MNRAS.460..497H, collision..2016MNRAS.463L..17A}. Their low surface density and strong field-star contamination make them challenging to identify, but the astrometric precision of \textit{Gaia} now enables systematic searches for these nearby, diffuse structures. The Hyades is an exceptional laboratory because its proximity permits a fine-grained comparison between observed tidal tails and dynamical models \citep{Hyadesintroduction..2011A&A...531A..92R, Gaia..2018A&A...616A..10G}.

Observational efforts initially focused on the Hyades core, using proper-motion catalogs and \textit{Hipparcos} astrometry to characterize membership within the tidal radius \citep{early_Hyades..1998A&A...331...81P}. \citet{Hyadesintroduction..2011A&A...531A..92R} extended the search using the PPMXL catalog, providing early evidence that the cluster extends beyond its classical tidal boundary. The extended tails became accessible with \textit{Gaia} \citep{Gaia..2016A&A...595A...1G, Gaia..2018A&A...616A..10G, 2019A&A...622L..13M}. In particular, \citet{roeser..2019A&A...621L...2R} mapped tidal-tail members in \textit{Gaia} DR2 using the convergent point (CP) method, establishing an empirical reference for the stream orientation near the cluster.

Numerical simulations are essential for connecting the present-day stream morphology to the underlying Galactic dynamics. Early $N$-body models primarily focused on reproducing the cluster's mass-loss rate and the evolution of its core properties within a static, axisymmetric Galactic potential \citep{Chumak..2005AstL...31..308C, Ernst..2011A&A...536A..64E}. The \textit{Gaia} detection of asymmetric tidal tails indicates that local dynamics are more complex than can be captured by an axisymmetric model \citep{J21..2021A&A...647A.137J}. Recent work has therefore begun to incorporate non-axisymmetric structure; for example, \citet{Tomas..2023A&A...678A.180T} explored the influence of a rotating bar and a flattened halo. However, the joint impact of the bar and spiral arms---and, in particular, how their pattern speeds map into observable stream diagnostics---remains insufficiently explored.

In this paper we perform a systematic survey of bar and spiral pattern speeds using high-resolution \texttt{PETAR} $N$-body simulations in a multi-component \texttt{AGAMA} Milky Way potential. We confront the models with \textit{Gaia} EDR3 using CP/CCP kinematic selection and an adaptive three-dimensional spatial matching procedure. We emphasize a single, consistently applied diagnostic across the manuscript: the stream orientation near the cluster and the longitudinal density distribution in a stream-aligned coordinate frame, including a prominent density peak at $Y_{\mathrm{rot}} \approx 0.1\,\mathrm{kpc}$. We find that models with $\Omega_b \simeq 40$--$45\,\mathrm{km\,s^{-1}\,kpc^{-1}}$ and $\Omega_s = 22.5\,\mathrm{km\,s^{-1}\,kpc^{-1}}$ best reproduce the observed density structure, supporting a relatively slow Galactic bar.

This paper is organized as follows. Section~\ref{sec:potential} describes the Galactic potential model and the $N$-body simulation setup. Section~\ref{sec:results} presents the simulated stream morphologies and kinematics across the explored $(\Omega_b,\Omega_s)$ parameter space. In Section~\ref{comparsion} we compare the simulations to observations, using the Röser CP-selected sample and \textit{Gaia}~EDR3 to constrain the pattern speeds via stream orientation and density structure. We discuss limitations and future observational tests in Section~\ref{discussion}.

\section{N-body simulations}

\subsection{Galactic potential} \label{sec:potential}

The Galactic gravitational potential was implemented using the \texttt{AGAMA} framework \citep{agama..2019MNRAS.482.1525V}. Our potential is based on the multi-component model described by \citet{Hunter..2024A&A...692A.216H}\footnote{The original potential configuration is available at \url{https://github.com/GalacticDynamics-Oxford/Agama/blob/master/py/example_mw_potential_hunter24.py}}, which includes contributions from the supermassive black hole Sgr~A$^\ast$, the nuclear star cluster, the nuclear stellar disk, the Galactic bar, the Galactic disk (comprising both axisymmetric and spiral arm components), and the dark matter halo. 

To account for the flattened dark matter distribution observed in recent studies \citep{Huang..2024NatAs...8.1294H,flatten_halo1..2019MNRAS.486.2995M, flatten_halo2..2024ApJ...967...89I}, we modified the original dark matter halo by replacing it with a triaxial spheroidal component. Following \citet{Tri_NFW..2016MNRAS.463.2623H}, this component is parameterized as a generalized NFW profile \citep{NFW..1997ApJ...490..493N} with the following configuration:
\begin{itemize}
    \item a scale radius $r_s = 14.39\,\mathrm{kpc}$;
    \item a vertical-to-radial axis ratio (flattening) $q = 0.82$;
    \item a scale density $\rho = 1.357 \times 10^7\,M_{\odot}\,\mathrm{kpc}^{-3}$;
    \item a profile shape defined by an inner slope $\gamma=1$, an outer slope $\beta=3$, and a transition sharpness $\alpha=1$.
\end{itemize}

\begin{figure}[h!]
    \centering
    \includegraphics[width=0.99\linewidth]{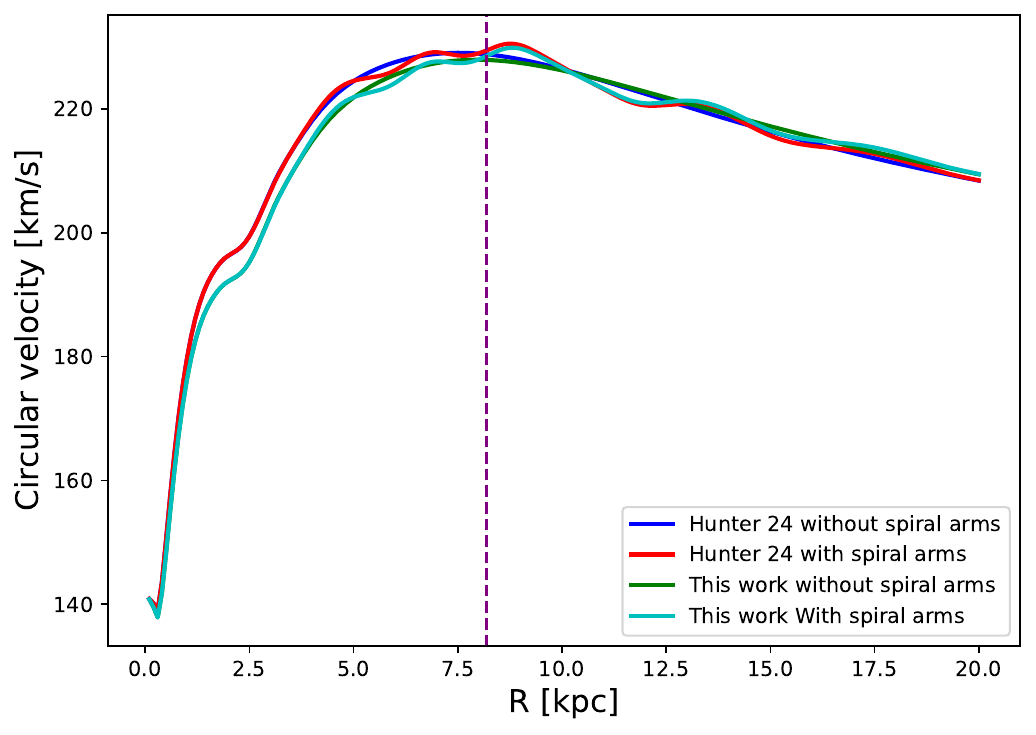}
    \caption{Circular-velocity curves: \citet{Hunter..2024A&A...692A.216H} without (blue) and with (red) spiral arms; potential model used in this study without (green) and with (cyan) spiral arms. Purple dot line represents R = 8.179kpc, which is close to the Sun's position used in this work}
    \label{fig:rotation curve}
\end{figure}

These parameters were specifically calibrated to ensure that the resulting circular velocity curve of our modified potential remains highly consistent with the original model described by \citet{Hunter..2024A&A...692A.216H} as shown in Fig.~\ref{fig:rotation curve}.

The dynamical impact of various resonance conditions is investigated through three distinct sets of potential configurations, generated by varying $\Omega_b$ and $\Omega_s$. In addition to these non-axisymmetric setups, we adopted the \texttt{MWPotential2014} model \citep{Galpy..2015ApJS..216...29B, WangMW2014..2021A&A...655A..71W, J21..2021A&A...647A.137J} via \texttt{galpy} as an axisymmetric baseline, representing a potential devoid of both bar and spiral arm perturbations. These models are summarized in Table~\ref{table:potential_sets}, providing a framework to systematically isolate the influence of each non-axisymmetric component.

\begin{table}[ht!]
\centering
\caption{Summary of Galactic potential configurations.}
\label{table:potential_sets}
\begin{tabular}{l c c}
\hline\hline
Set & $\Omega_b$ & $\Omega_s$ \\
\hline
Set 1 (Bar only)        & $\Omega_{\mathrm{grid}}$ & --- \\
Set 2 (Bar + Fixed arms) & $\Omega_{\mathrm{grid}}$ & $22.5$ \\
Set 3 (Fixed bar + Arms) & $37.5$ & $\Omega_{\mathrm{grid}}$ \\
Baseline (Axisymmetric)  & \multicolumn{2}{c}{\texttt{MWPotential2014}} \\
\hline
\multicolumn{3}{p{0.95\columnwidth}}{\small \textbf{Notes.} All pattern speeds are in units of $\mathrm{km\,s^{-1}\,kpc^{-1}}$. The symbol $\Omega_{\mathrm{grid}}$ denotes the set of discrete pattern speeds: 10, 20, 25, 29, 35, 37.5, 40, 42.5, 45, 50, 55.} \\
\end{tabular}
\end{table}

The simulation reference frame is anchored by a Sun-Galactic center distance of $R_0 = 8.179\,\mathrm{kpc}$ and a solar circular velocity of $V_c(R_0) = 229\,\mathrm{km\,s^{-1}}$ \citep{SUn_distance..2022A&A...657L..12G, Hunter..2024A&A...692A.216H}, ensuring consistency with local kinematic constraints. Furthermore, the solar peculiar motion is taken from \citet{solor_motion..2010MNRAS.403.1829S}, with $(U_{\odot}, V_{\odot}, W_{\odot}) = (11.1, 12.24, 7.25)\,\mathrm{km\,s^{-1}}$. 

\subsection{Simulation setup}

In this work, we employed the high-performance $N$-body code \texttt{PETAR} \citep{Petar..2020MNRAS.497..536W} to simulate the dynamical evolution of the Hyades \citep{WangMW2014..2021A&A...655A..71W}. \texttt{PETAR} has the ability to efficiently handle both the internal collisional dynamics of the star cluster and the external Galactic tidal field. Specifically, \texttt{PETAR} employs a hybrid algorithm that effectively accounts for gravitational interactions across all spatial scales by combining dedicated methods for short-, middle-, and long-range forces using \texttt{sdar} and \texttt{fdps} \citep{Petar_SDAR..2020MNRAS.491.2413W, Petar_Hermite..2003gnbs.book.....A, Petar_tree_algor..1986Natur.324..446B, 69-Iwasawa.2016, 70-Iwasawa.2020,71-Namekata.2018}. This multi-scale treatment ensures the high numerical precision and computational efficiency required to simulate the long-term evolution and mass loss of stellar systems over several hundred megayears. 

The present-day parameters adopted for the Hyades cluster in our simulations were partially based on \cite{WangMW2014..2021A&A...655A..71W}, as shown in Table \ref{table:present-day-parameters}. Then we used the \texttt{Astropy SkyCoord} package to transform the present-day parameters into Galactocentric positions and velocities \citep{astropy2013..2013A&A...558A..33A, astropy2018..2018AJ....156..123A, astropy2022..2022ApJ...935..167A}.

To simulate the dynamical evolution of the Hyades stream, we first determined the initial phase-space coordinates of the cluster 648\,Myr ago \citep{WangMW2014..2021A&A...655A..71W}. For each gravitational potential model, we performed a backward integration of a test particle starting from the cluster's current Galactocentric position. However, due to internal dynamical interactions and centroid shifts inherent in $N$-body systems, the final centroid of the simulated cluster typically deviates from the observed present-day position, even if a test particle returns to its origin. To account for this discrepancy, we used the initial conditions derived from the test-particle integration as a first guess for a Nelder-Mead optimization \citep{Nelder_mead..10.1093/comjnl/7.4.308}. We iteratively refined these initial coordinates until the simulated cluster's centroid at the end of the 648\,Myr evolution satisfied the following convergence criteria: the positional differences in $x$ and $y$ were less than $100\,\mathrm{pc}$, the $z$ difference was less than $20\,\mathrm{pc}$, the velocity magnitude deviation was below $10\,\mathrm{km\,s^{-1}}$, and the angular misalignment between the simulated and observed velocity vectors was within $5^\circ$.

\begin{table}[t!]
\centering
\caption{Present-day parameters of the Hyades (in the ICRS frame)}
\begin{tabular}{|l|c|}
\hline
\textbf{Parameter} & \textbf{Value} \\
\hline
RA [deg] & 67.985 \\
Dec [deg] & 17.012 \\
Distance [pc] & 47.501 \\
pm(RA, cos(Dec)) [mas/yr] & 101.005 \\
pm(Dec) [mas/yr] & $-28.490$ \\
Radial velocity [km/s] & 39.96 \\
Age [Myr] & 648 \\
\hline
\end{tabular}
\label{table:present-day-parameters}
\end{table}

We employed the \texttt{McLuster} package \citep{mcluster..2011MNRAS.417.2300K,Wang2019} to construct the initial model of the Hyades cluster. The cluster was parameterized as a Plummer sphere with an initial total mass of $800\,M_{\odot}$ and an initial half-mass radius of $r_h = 1\,\mathrm{pc}$. Individual stellar masses were sampled from a \citet{KroupaIMF..2001MNRAS.322..231K} initial mass function (IMF). To account for stellar evolution, we use the \textsc{sse} and \textsc{bse} modules \citep{Hurley2000,Hurley2002} updated by \cite{Banerjee2020}, assuming a solar metallicity of $Z = 0.02$ throughout the simulation.

\subsection{Simulation Results} \label{sec:results}

\subsubsection{Stream Morphology}

We first summarize the qualitative dependence of the Hyades stream morphology on $(\Omega_b,\Omega_s)$, and we then use the orientation- and density-based observational comparisons in Section~\ref{comparsion} to identify the subset of models consistent with \textit{Gaia}.

We first explored the parameter space of $\Omega_b$ in the absence of spiral arms. Throughout these simulations, the bar length was kept fixed to ensure that the observed morphological changes resulted solely from the variation in rotation speed.

Two critical $\Omega_b$ associated with orbital resonances are of particular interest: the corotation resonance (CR) and the outer Lindblad resonance (OLR). 

The CR occurs at the radius where $\Omega_b$ matches the local circular orbital frequency, while the OLR represents a higher-frequency resonance where the bar potential couples with the radial epicyclic motion of the stars. Assuming the guiding radius of the Hyades is approximately solar, $\Omega_b$ required for CR is estimated as $\Omega_{\mathrm{CR}} = V_c / R_{\mathrm{CR}} \simeq 29\,\mathrm{km\,s^{-1}\,kpc^{-1}}$. Furthermore, the frequency associated with the OLR can be estimated following \citet{OLR..2008gady.book.....B, OLR..2016MNRAS.460..497H}:

\begin{equation}
    \Omega_{\mathrm{OLR}} = \Omega_{\phi} + \frac{N}{2}\Omega_{R},
\label{OLR}
\end{equation}

where $\Omega_{\phi}$ and $\Omega_{R}$ are the azimuthal and radial frequencies, respectively. By setting $N=1$, we obtain an OLR pattern speed of $\Omega_{\mathrm{OLR}} \simeq 50\,\mathrm{km\,s^{-1}\,kpc^{-1}}$.

\begin{figure*}[ht!]
    \centering
    \includegraphics[width=0.99\linewidth]{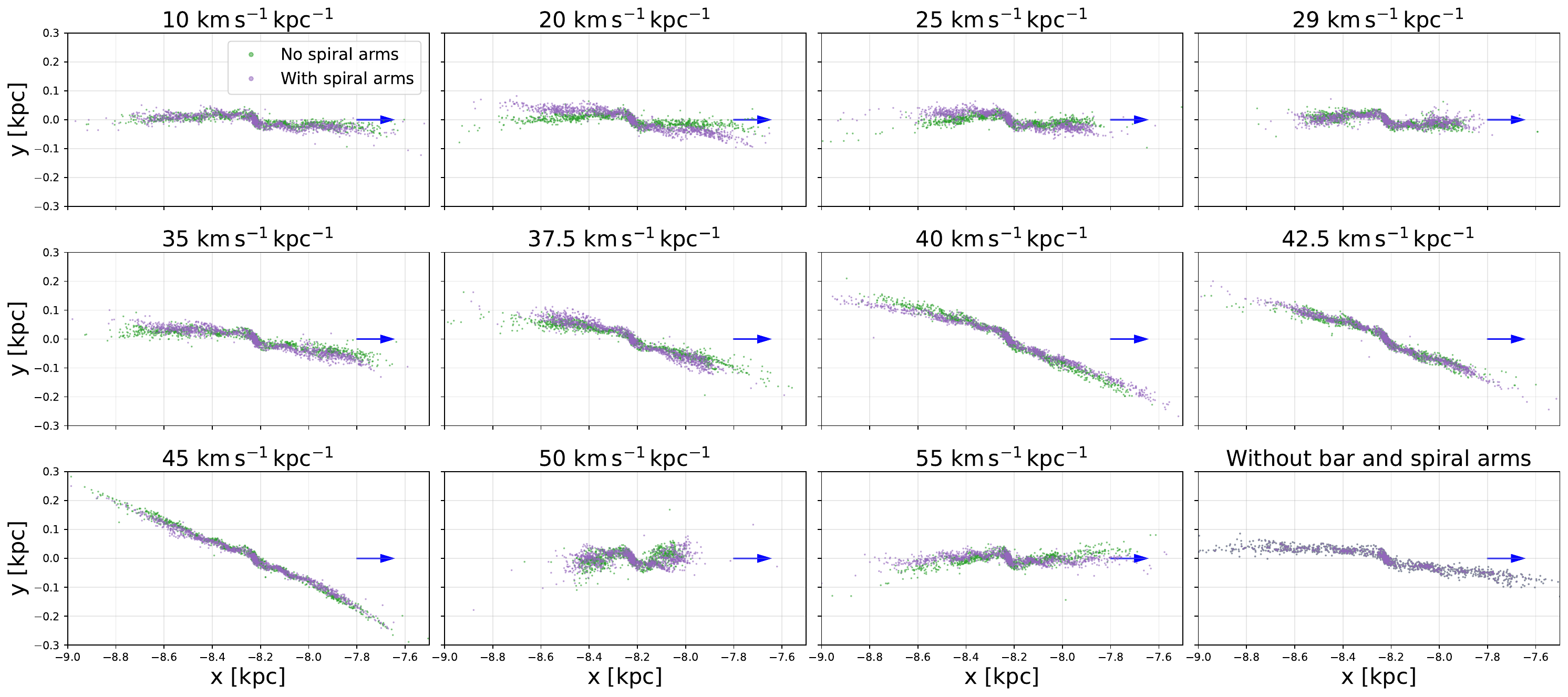}
    \caption{Galactocentric positions of stars in Hyades simulations with bar pattern speeds $\Omega_b$ ranging from $10$ to $55\,\mathrm{km\,s^{-1}\,kpc^{-1}}$. Models including spiral arms are shown in purple, with $\Omega_s$ fixed at $22.5\,\mathrm{km\,s^{-1}\,kpc^{-1}}$; the corresponding bar-only models are shown in green. After shifting and rotating each model into a stream-aligned frame, the blue arrow indicates the centroid velocity direction adopted as the $x$-axis.}
    \label{fig:diff_bar}
\end{figure*}

The Galactocentric positions of the simulated Hyades stream are presented in Fig.~\ref{fig:diff_bar}. 
Upon completion of each simulation, the simulated streams were spatially aligned with the observed counterpart by applying a positional shift. The velocity direction of the stream's centroid was then adopted as the $x$-axis, corresponding to the direction indicated by the blue arrow in the accompanying figure. With the centroid position held fixed, the positions and velocities of all stream stars were rotated into this new coordinate frame.

In the low-speed regime, where $\Omega_b$ is below $\Omega_{\mathrm{CR}}$, the angle between the stellar stream and its centroid velocity vector measured in the inertial Galactocentric frame (hereafter referred to as the misalignment angle) remains essentially zero and exhibits minimal sensitivity to increases in $\Omega_b$. However, as $\Omega_b$ approaches the value at the corotation radius, the stream length undergoes a modest reduction due to corotation effects. 

As $\Omega_b$ continues to increase beyond the corotation radius, the misalignment angle gradually becomes more significant. In contrast, once $\Omega_b$ exceeds $\Omega_{\mathrm{OLR}}$, the misalignment angle converges back toward the configuration observed at low $\Omega_b$, maintaining values near zero with little variation. Notably, the reduction in stream length near the OLR is significantly more pronounced than the modest decrease observed at the corotation radius. 

In all simulations, the final positions of the stream's centroid and the direction of its centroid velocity align well with the observed properties of the Hyades stream. Consequently, the variations in both the misalignment angle and the length of the stream reflect how the bar's rotating potential reshapes the spatial orientation and extent of the tidal tails. Our findings are consistent with the simulation results of \citet{Tomas..2023A&A...678A.180T}. 

Subsequently, we included the spiral arms component in the potential, fixing $\Omega_s$ at \(22.5\,\mathrm{km\,s^{-1}\,kpc^{-1}}\), and again varied $\Omega_b$. We found that the resulting configurations were remarkably similar to those obtained in the bar-only simulations (Fig.~\ref{fig:diff_bar}), indicating that the influence of the spiral arms on the Hyades is secondary to that of the bar. This result is expected, as at the Galactocentric distance of the Hyades, the magnitude of the gravitational potential contributed by the spiral arms is only approximately 1\% of that contributed by the bar \citep{Hunter..2024A&A...692A.216H}. 

However, we note that for bar pattern speeds of $20$ and $25\,\mathrm{km\,s^{-1}\,kpc^{-1}}$, the inclusion of spiral arms leads to a more discernible change in the stream orientation compared to the bar-only case. This phenomenon likely arises from a resonant interaction between the bar and the spiral arms, as their respective pattern speeds in these specific models are close to each other, enhancing the total non-axisymmetric perturbation experienced by the cluster. Nevertheless, as we show in Section~\ref{comparsion}, varying $\Omega_s$ can still modulate the fine-grained density structure along the stream even when the global orientation is largely bar-driven.

Furthermore, we compared these results with the baseline model (see Table~\ref{table:potential_sets}). In this axisymmetric case, the misalignment angle remains relatively small, further confirming that the significant deviations observed in our other simulation sets are primarily driven by the non-axisymmetric bar and spiral arms potential.

% maybe discuss the slightly difference in 20,25?

\begin{figure*}[ht!]
    \centering
    \includegraphics[width=0.99\linewidth]{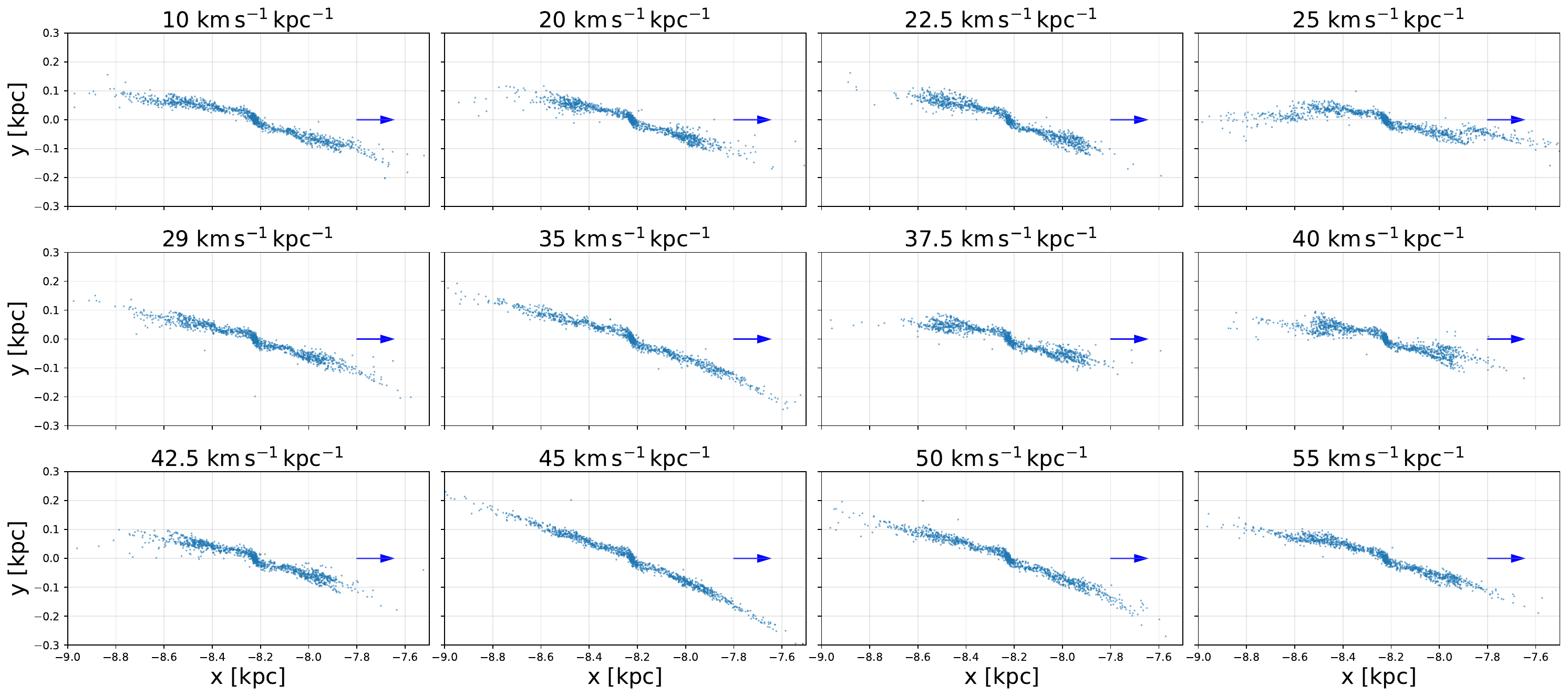}
    \caption{Similar to Figure~\ref{fig:diff_bar}, but varying the spiral pattern speed $\Omega_s$ from $10$ to $55\,\mathrm{km\,s^{-1}\,kpc^{-1}}$ with $\Omega_b$ fixed at $37.5\,\mathrm{km\,s^{-1}\,kpc^{-1}}$.}
    \label{fig:diff_spiral}
\end{figure*}

We then fixed $\Omega_b$ at $37.5\,\mathrm{km\,s^{-1}\,kpc^{-1}}$, a value consistent with recent observational estimates, and varied $\Omega_s$ \citep{Hunter..2024A&A...692A.216H}. The results are presented in Fig.~\ref{fig:diff_spiral}. Overall, the misalignment angle does not exhibit significant sensitivity to changes in $\Omega_s$; however, both the total length of the stellar stream and the emergence of internal substructures show discernible variations across the different models.

\subsubsection{Misalignment angle}

\begin{figure*}[ht!]
\centering
\gridline{\fig{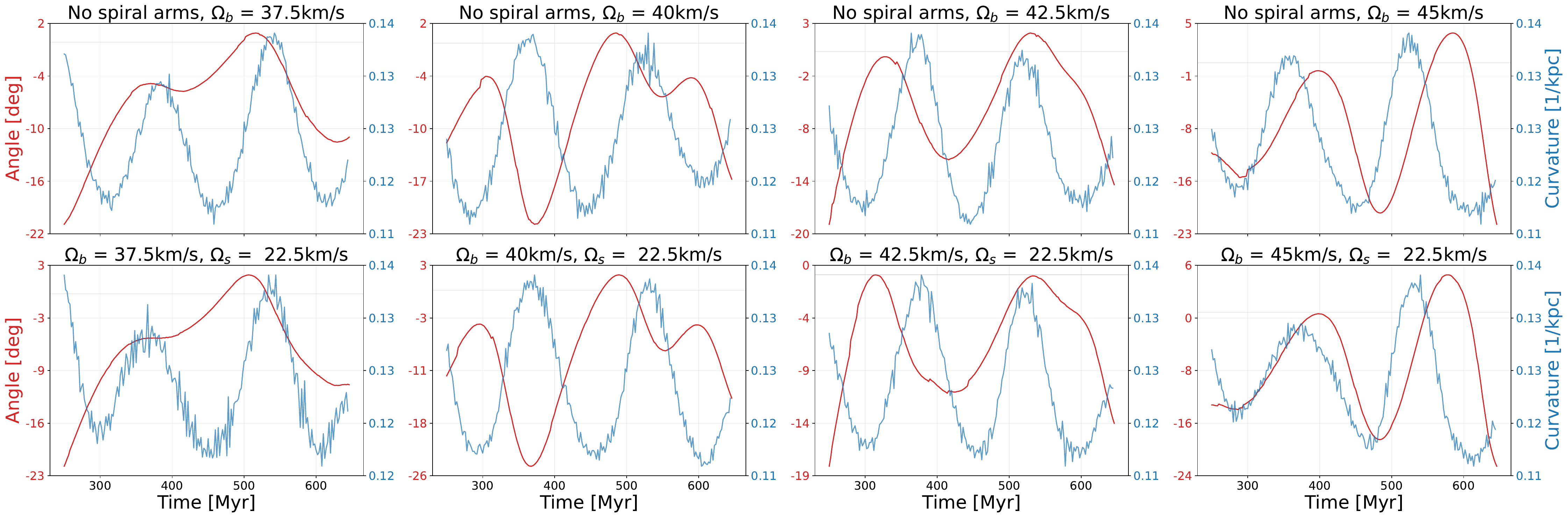}{0.99\textwidth}{(a)}}
\gridline{\fig{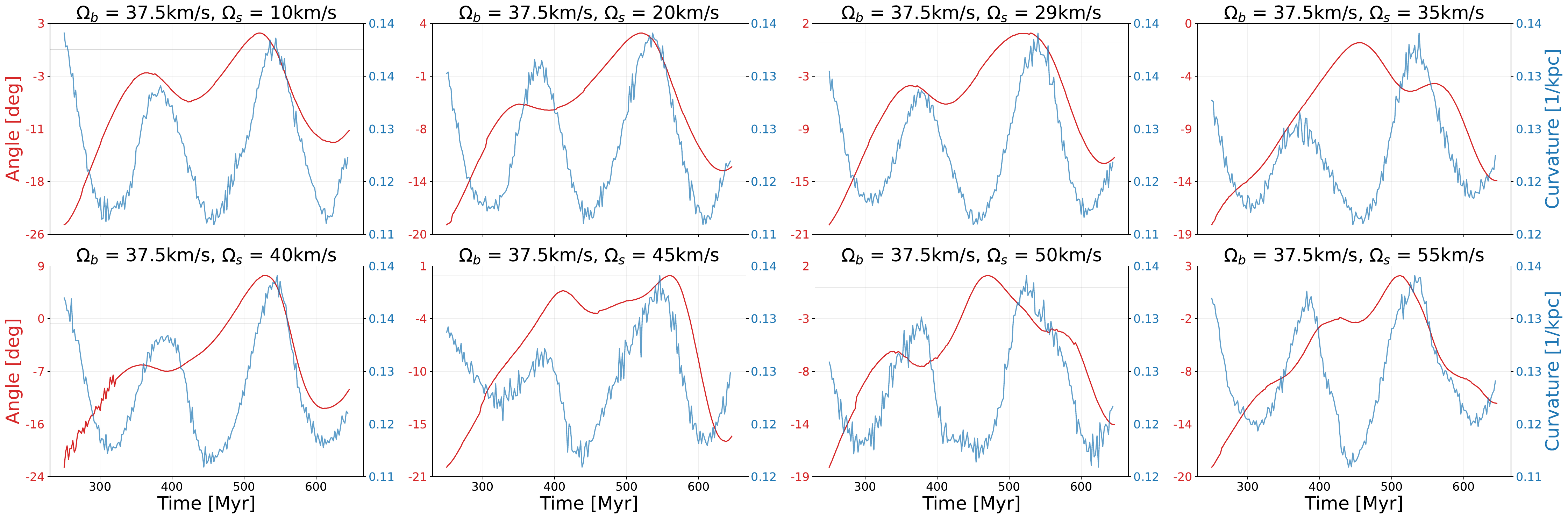}{0.99\textwidth}{(b)}}
\caption{Time evolution of the misalignment angle (red solid lines) and the local orbital curvature (blue solid lines) for the spatially matched models. The $x$-axis represents the time since the cluster formation in Myr. The left $y$-axis indicates the misalignment angle in degrees, and the right $y$-axis shows the orbital curvature in units of $\mathrm{kpc}^{-1}$. 
Panel (a) shows the impact of $\Omega_b$ (with and without spiral arms), and Panel (b) focuses on the impact of $\Omega_s$ at a constant $\Omega_b$.
\label{fig:misaligned_angel}}
\end{figure*}

%By comparing the spatial distribution in the XY plane of the Galactocentric coordinate system with the Röser dataset, we filtered out a subset of models (hereafter spatial matched models). Compared to other models, an evident characteristic of
We further investigate the mechanism driving the misalignment angles in several models. Ideally, under a static and symmetric gravitational potential, Hyades should not exhibit significant misalignment angles. Therefore, the emergence of misalignment angles is more likely attributable to an asymmetric and time-varying gravitational potential, particularly due to the bar component. This is also evident in Fig.~\ref{fig:diff_bar}, where $\Omega_b$ significantly influences the misalignment angles of Hyades at the present observational epoch after its evolution. 

To investigate this phenomenon more thoroughly, for each spatially matched model (see Section~\ref{comparsion} for definition), we computed the time evolution of the misalignment angle after the formation of the stellar stream, as shown in Fig.~\ref{fig:misaligned_angel}. Clearly, different $\Omega_b$ and $\Omega_s$ have a substantial impact on the evolution of the misalignment angle. Moreover, certain modulation patterns can be distinctly observed in this evolution, and the modulation period closely matches the orbital period of Hyades around the Galactic center. Accordingly, the evolution of curvature over time is also plotted with blue curves in Fig.~\ref{fig:misaligned_angel}. 

Here, the ``curvature'' refers to the geometric curvature of the three-dimensional centroid trajectory of the stellar stream as it evolves over time. To compute this numerically from our simulation data, we treat the centroid position as a parametric curve $\mathbf{r}(t)$ in 3D space. We then estimate the first and second-order derivatives, $\mathbf{r}'(t)$ and $\mathbf{r}''(t)$, using finite difference schemes. The curvature $\kappa$ at each epoch $i$ is determined according to the standard differential geometry definition:
\begin{equation}
    \kappa_i = \frac{\| \mathbf{r}'_i \times \mathbf{r}''_i \|}{\| \mathbf{r}'_i \|^3 + \epsilon},
\end{equation}
where $\epsilon$ is a small constant ($10^{-16}$) included to ensure numerical stability.

In some models (e.g., $\Omega_b=45\,\mathrm{km\,s^{-1}\,kpc^{-1}}$ and $42.5\,\mathrm{km\,s^{-1}\,kpc^{-1}}$), a clear modulation relationship between the misalignment angle and curvature is evident, whereas in other models this relationship is less pronounced. As a result, the observed correlation may suggest a loose link between the evolution of the misalignment angle and the orbital curvature. This would imply that the time-varying tidal forces experienced by the cluster at specific orbital positions could act as one of the contributing factors—alongside other complex orbital dynamics—in shaping the stream's orientation.

\subsubsection{Proper motions}

In addition to the spatial distribution, the kinematic properties of the various models merit examination. We calculated the proper motion distributions across all simulations. We found that while different models exhibit morphological variations in space, the kinematics of the stream core remain broadly similar. Discrepancies primarily emerge in the wings of the proper motion distributions, which correspond to the stars located in the distal tidal tails.

\begin{figure}[ht!]
    \centering
    \includegraphics[width=0.99\linewidth]{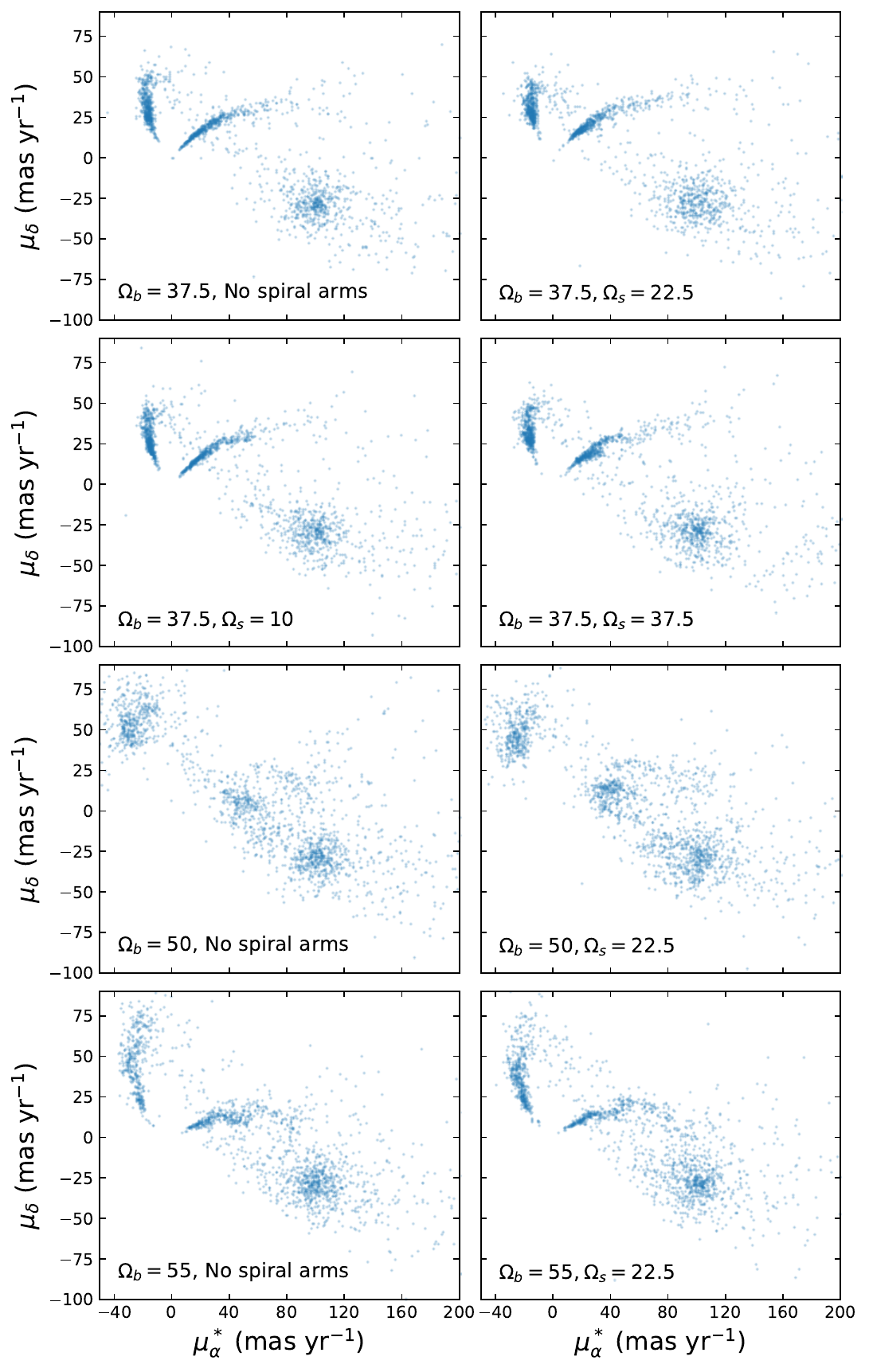}
    \caption{Proper motion distributions for a representative selection of models. The first three rows contrast the kinematics with and without spiral arms ($\Omega_s = 22.5\,\mathrm{km\,s^{-1}\,kpc^{-1}}$) at bar pattern speeds of $\Omega_b = 37.5$, $50$, and $55\,\mathrm{km\,s^{-1}\,kpc^{-1}}$, respectively. The bottom row focuses on the impact of different spiral arm speeds ($\Omega_s = 10$, $37.5\,\mathrm{km\,s^{-1}\,kpc^{-1}}$) for a constant bar speed of $\Omega_b = 37.5\,\mathrm{km\,s^{-1}\,kpc^{-1}}$.}
    \label{fig:proper_motion}
\end{figure}

To clearly illustrate these kinematic behaviors, Figure~\ref{fig:proper_motion} presents the 2D proper motion diagrams for a representative subset of our models, arranged in pairwise comparisons. The first and fourth rows demonstrate the impact of spiral arms. These panels represent the dynamical behavior seen in the majority of our models (not plotted here), where the presence or varying $\Omega_s$ introduce only minor perturbations to the overall proper motion distribution. This implies that, under most standard configurations ($\Omega_b < \Omega_{\mathrm{OLR}}$), the proper motion of the stream is not highly sensitive to the bar and spiral arms.

In contrast, the second and third rows display the extreme cases with fast bars ($\Omega_b = 50$ and $55\,\mathrm{km\,s^{-1}\,kpc^{-1}}$, respectively). Under these conditions, the proper motion distributions of the stream tails deviate significantly from the typical models. The distinct behavior at $\Omega_b = 50\,\mathrm{km\,s^{-1}\,kpc^{-1}}$ is physically expected. As previously shown in Fig.~\ref{fig:diff_bar}, this pattern speed coincides with $\Omega_{\mathrm{OLR}}$ which strongly truncates the stream's spatial extent. Furthermore, as $\Omega_b$ increases beyond $\Omega_{\mathrm{OLR}}$, the proper motion of the distal tail progressively diverges from the trajectory established by slower bars.

Nevertheless, despite these dramatic kinematic variations in the tidal tails, the dense core of the cluster consistently occupies a common locus in the proper motion space across all models. This sensitivity pattern confirms that the adopted Galactic potential primarily governs the velocity-space evolution of the outer stream regions, while the core kinematics remain highly robust.

\section{Observational Constraints on the Bar and Spiral Arms}\label{comparsion}

\subsection{Spatial Distribution}

We constrain $(\Omega_b,\Omega_s)$ in two steps. First, we require the simulated stream orientation near the cluster to match the empirical direction traced by the Röser CP-selected sample. Second, for the orientation-selected models we use \textit{Gaia} EDR3 to compare the internal density structure along the stream, focusing on the location and prominence of the $Y_{\mathrm{rot}} \approx 0.1\,\mathrm{kpc}$ overdensity that drives our final pattern-speed constraint.

Identifying member stars in the distant tidal tails of the Hyades stream remains challenging because of strong field-star contamination and the low surface density of the stream \citep{Hyadesintroduction..2011A&A...531A..92R}. \citet{roeser..2019A&A...621L...2R} identified a relatively clean set of candidates near the cluster using the convergent point (CP) method applied to \textit{Gaia} DR2 (hereafter the Röser dataset), which does not assume a specific $N$-body model or Galactic potential model \citep{CP..2009A&A...497..209V}. Although the Röser dataset is concentrated near the cluster, one tail segment is visible in the Galactocentric $XY$ plane, allowing us to infer the local stream direction and reject simulation models whose projected orientation is inconsistent with the data.

\begin{figure}[h!]
    \centering
    \includegraphics[width=0.99\linewidth]{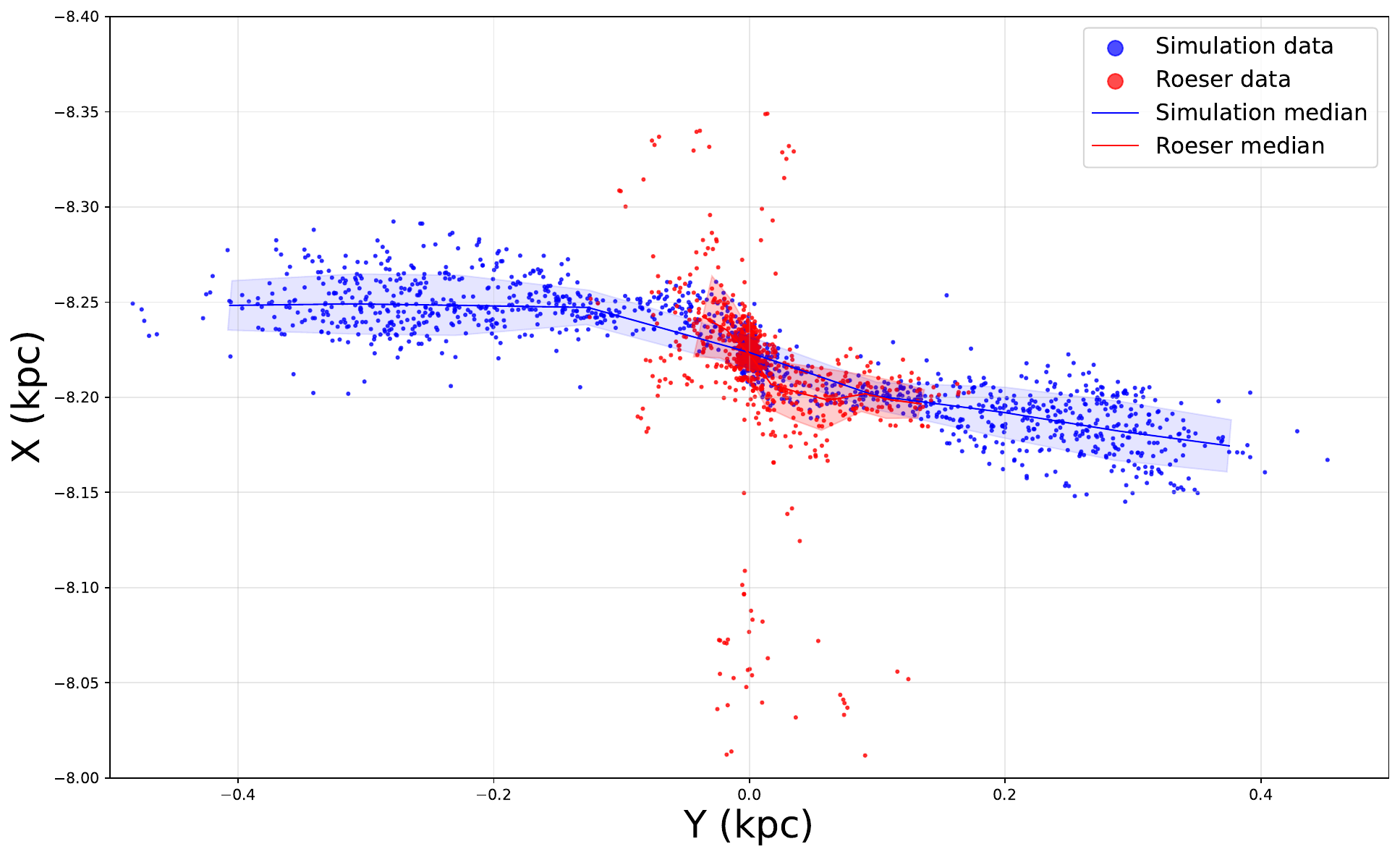}
    \caption{Galactocentric positions of the Hyades stream from a simulation with $\Omega_b = 37.5\,\mathrm{km\,s^{-1}\,kpc^{-1}}$ and $\Omega_s = 22.5\,\mathrm{km\,s^{-1}\,kpc^{-1}}$ (blue) and from the Röser dataset (red). The solid lines in corresponding colors show the PCA-derived stream directions, with the shaded regions representing the 1$\sigma$ dispersion of the stream.}
    \label{fig:median}
\end{figure}

We employed Principal Component Analysis (PCA) to determine the principal orientation of the Hyades stream in both the Röser dataset and each simulated model, characterizing the stream direction with a median line \citep{PCA..2016RSPTA.37450202J}. To mitigate the impact of spatial scatter in the $XY$ plane and ensure a robust directional fit, we restricted the PCA calculation to stars within the $X$-coordinate range of $[-8.3, -8.15]$\,kpc. Fig.~\ref{fig:median} illustrates a comparison between the Röser dataset and a representative simulation with $\Omega_b = 37.5\,\mathrm{km\,s^{-1}\,kpc^{-1}}$ and $\Omega_s = 22.5\,\mathrm{km\,s^{-1}\,kpc^{-1}}$.

\begin{figure}[h!]
    \centering
    \includegraphics[width=0.9\linewidth]{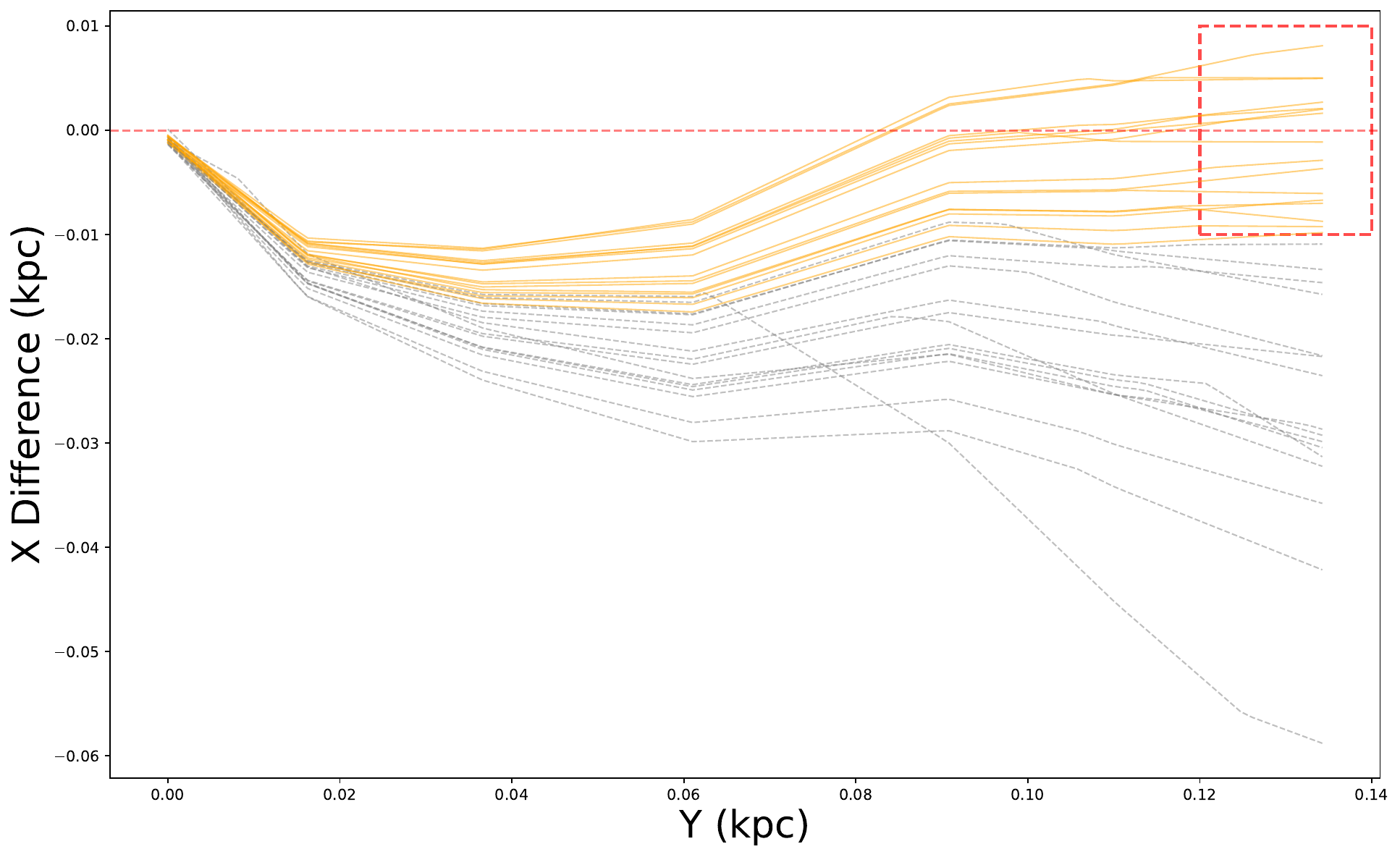}
    \caption{X-directional offset between the median lines of each model and the Röser dataset. The red dashed rectangle highlights the region defined by a Y-coordinate of 0.12–0.14 kpc and an X-offset within $\pm$ 0.01 kpc. Models satisfying these criteria (orange tracks) include: (i) models with $\Omega_b$ of $37.5, 40, 42.5,$ and $45\,\mathrm{km\,s^{-1}\,kpc^{-1}}$ (both without spiral arms and with a $\Omega_s = 22.5\,\mathrm{km\,s^{-1}\,kpc^{-1}}$ spiral arm component), and (ii) models with a fixed $\Omega_b$ of $37.5\,\mathrm{km\,s^{-1}\,kpc^{-1}}$ and specific $\Omega_s \in \{10, 20, 29, 35, 40, 45, 50, 55\}\,\mathrm{km\,s^{-1}\,kpc^{-1}}$. All other models are depicted with gray dashed lines.}
    \label{fig:median_difference}
\end{figure}

For each model, we computed the offset in the $X$-direction between the median lines of the simulated stream and the Röser dataset as a function of the $Y$-coordinate. Given that the Röser dataset primarily populates one side of the stream over a limited longitudinal extent, this comparison was restricted to $Y > 0$\,kpc, reaching the farthest detected point in the observations. The results are illustrated in Fig.~\ref{fig:median_difference}. Based on these offsets, we selected candidate models where the $X$-offset remained below $0.01$\,kpc (corresponding approximately to the $1\sigma$ dispersion of the stream as illustrated by the shaded region in Fig.~\ref{fig:median}) within the range $Y \in [0.12, 0.14]$\,kpc (where the Röser dataset exhibits both a smaller dispersion and a more stable, well-defined orientation). 

We find that for $\Omega_b$ between $37.5$ and $45\,\mathrm{km\,s^{-1}\,kpc^{-1}}$, the simulated stream directions consistently match the Röser dataset well (hereafter spatially matched models), irrespective of the presence of spiral arms. This result is in good agreement with recent estimates of $\Omega_b$ \citep{barspeed..2019MNRAS.489.3519C, Hunter..2024A&A...692A.216H}. Furthermore, nearly all models with varying $\Omega_s$ also exhibit close agreement with the Röser dataset. This is expected, as $\Omega_b$ was fixed at $37.5\,\mathrm{km\,s^{-1}\,kpc^{-1}}$ in these experiments, further corroborating that $\Omega_s$ has a negligible impact on the overall orientation of the stream.

\subsection{Convergent Point Method}

% CP and CCP
We then proceeded to investigate the properties of the spatially matched models in velocity space using the CP method. This approach considers that star cluster members move as a single entity with a common three-dimensional space velocity vector. Its projection onto the celestial sphere yields a convergent point. By adopting the bulk space velocity vector of the cluster, the CP method predicts the expected transverse motion of a member star at any given position on the sky. This predicted motion can be decomposed into two orthogonal components: one directed toward the convergent point ($V_{\parallel, \mathrm{pred}}$) and one perpendicular to it ($V_{\perp, \mathrm{pred}}$). For a star perfectly sharing the cluster's common space velocity, the perpendicular component $V_{\perp, \mathrm{pred}}$ is by definition zero, while $V_{\parallel, \mathrm{pred}}$ represents the full magnitude of the projected transverse velocity. By calculating these two velocity components, it is possible to identify cluster members. The CP method is particularly effective for nearby clusters whose members span large areas on the sky. 

\begin{figure*}[ht!]
    \centering
    \includegraphics[width=0.99\linewidth]{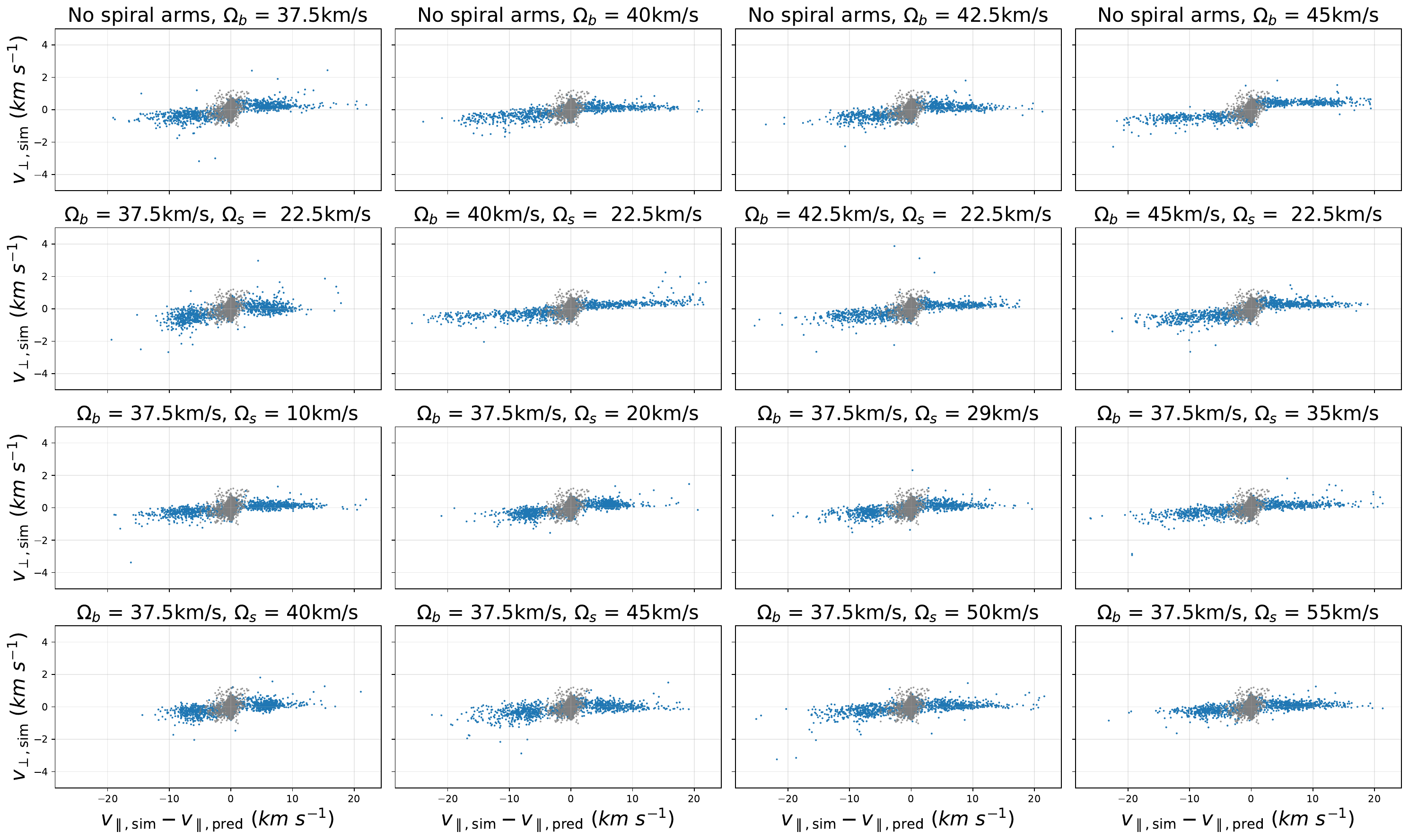}
    \caption{CP diagram of spatially matched models (blue points) and Röser dataset (grey points). X-axis: Difference between the simulated/observed velocity towards the convergent point and the expected velocity calculated using the convergent point method Y-axis: Velocity component perpendicular to the convergent point direction from simulations/observations.}
    \label{fig:CP}
\end{figure*}

For all spatially matched models, we employed the CP method to calculate the difference between the transverse motion of each star within the simulated stellar stream and the expected transverse motion, presenting the results in CP diagram as shown in Fig.~\ref{fig:CP}. The CP diagrams exhibit notable variations in length and dispersion across different models. For comparative purposes, we have overplotted data from the Röser dataset as gray points in the same figures. 
We note that although some field contaminants may still be present in the observational sample, we did not further clean it using \textit{Gaia} line-of-sight velocities. This is because the limited availability and relatively high uncertainties of RVS data for fainter stars would significantly compromise the completeness of the sample. 
It can be observed that all models demonstrate similarity to the Röser dataset in the central regions, providing further confirmation that while the extended tidal tails show significant variations among different models, the spatial and kinematic properties in the core regions of the stellar streams remain comparable.

\subsection{Compact Convergent Point Method}

\begin{figure*}[ht!]
    \centering
    \includegraphics[width=0.99\linewidth]{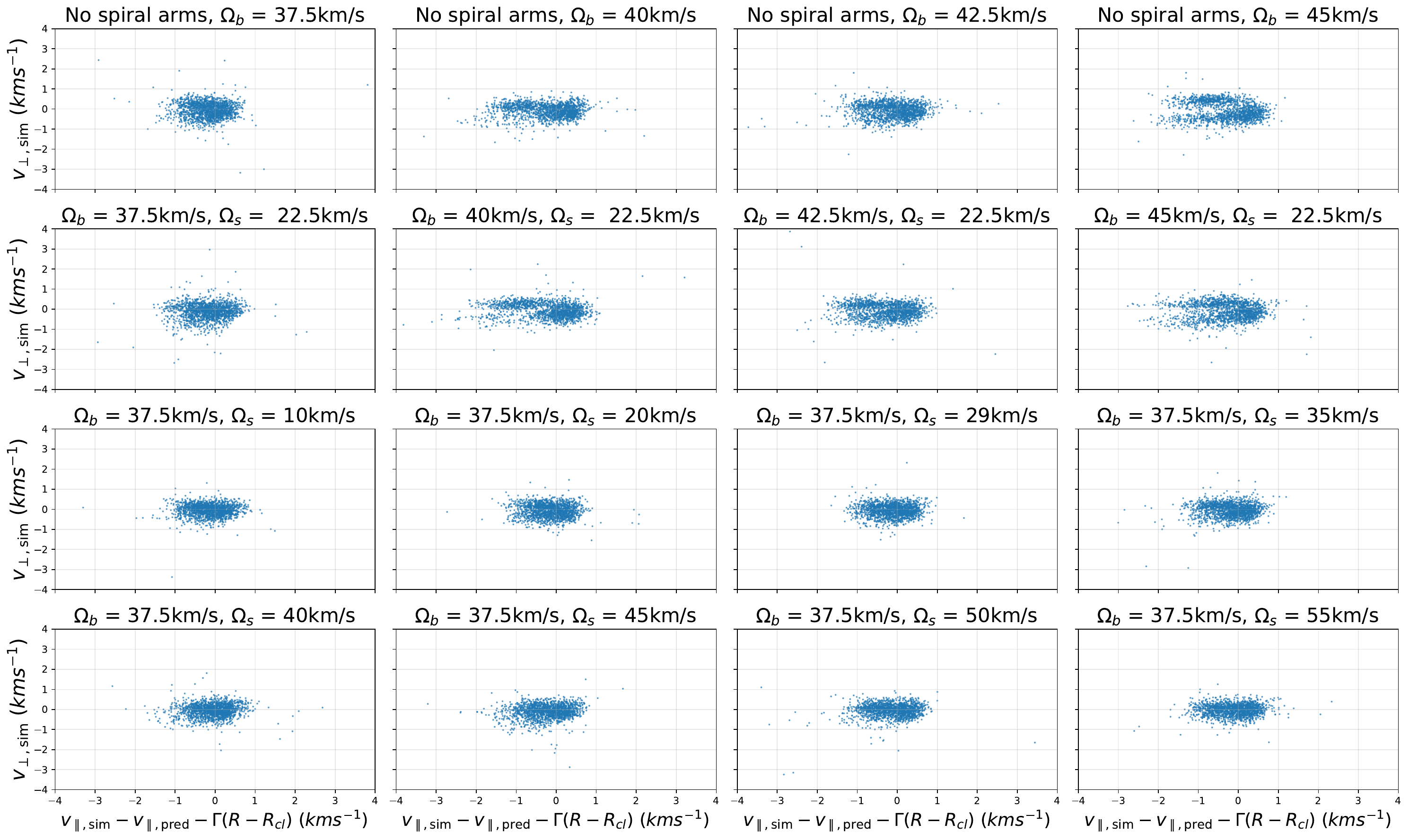}
    \caption{Same as Figure \ref{fig:CP}, but with the CCP method applied to remove the linear dependence in the velocity along the convergent point direction. $\Gamma$ is the slope of linear relationship between $V_{||,sim}-V_{||,pred}$ and $R-R_{cl}$.}
    \label{fig:CCP}
\end{figure*}

To address the extended structures in the tails of the CP diagram, \citet{J21..2021A&A...647A.137J} proposed a compact convergent point (CCP) method. This approach compresses the extended structures in the CP diagram into smaller, more concentrated regions by correcting for the linear relationship between $V_{||,sim}-V_{||,pred}$ and the distance of stream stars from the cluster center ($R-R_{cl}$). We have identified this relationship across all spatially matched models and have consequently adopted the CCP method to refine the CP diagram presented in Fig.~\ref{fig:CP}. The resulting diagram is shown in Fig.~\ref{fig:CCP}. After correction with the CCP method, all CP diagrams exhibit clustered concentrations around the (0,0) position. However, several models, such as the one without spiral arms and with $\Omega_b = 22.5\,\mathrm{km\,s^{-1}\,kpc^{-1}}$, display some substructures. 

% discuss the substructures in CCP diagram

Based on the results of the CCP method, we can further identify potential member stars from observations for each spatially matched model. We utilize data from \textit{Gaia} Early Data Release 3 \citep{Gaia_eDR3..2021A&A...649A...1G}. We retrieved the \textit{Gaia} data from the ESA archive\footnote{\url{https://gea.esac.esa.int/archive/}} using an ADQL query identical to the approach of \cite{Tomas..2023A&A...678A.180T}. The selection criteria include a parallax threshold of $\varpi \geq 1.0$\,mas, a parallax signal-to-noise ratio ($\varpi/\sigma_\varpi$) of at least 10, and a Renormalised Unit Weight Error (RUWE) below 1.4 to ensure reliable astrometric solutions. This query results in a dataset of 38,256,266 objects. Furthermore, we restricted our sample to sources with available extinction estimates ($A_G$ and $E(BP-RP)$) provided by the GSP-Phot Aeneas pipeline \cite{extinction_selection_2023A&A...674A..31D}. 

\begin{figure}[h!]
    \centering
    \includegraphics[width=0.9\linewidth]{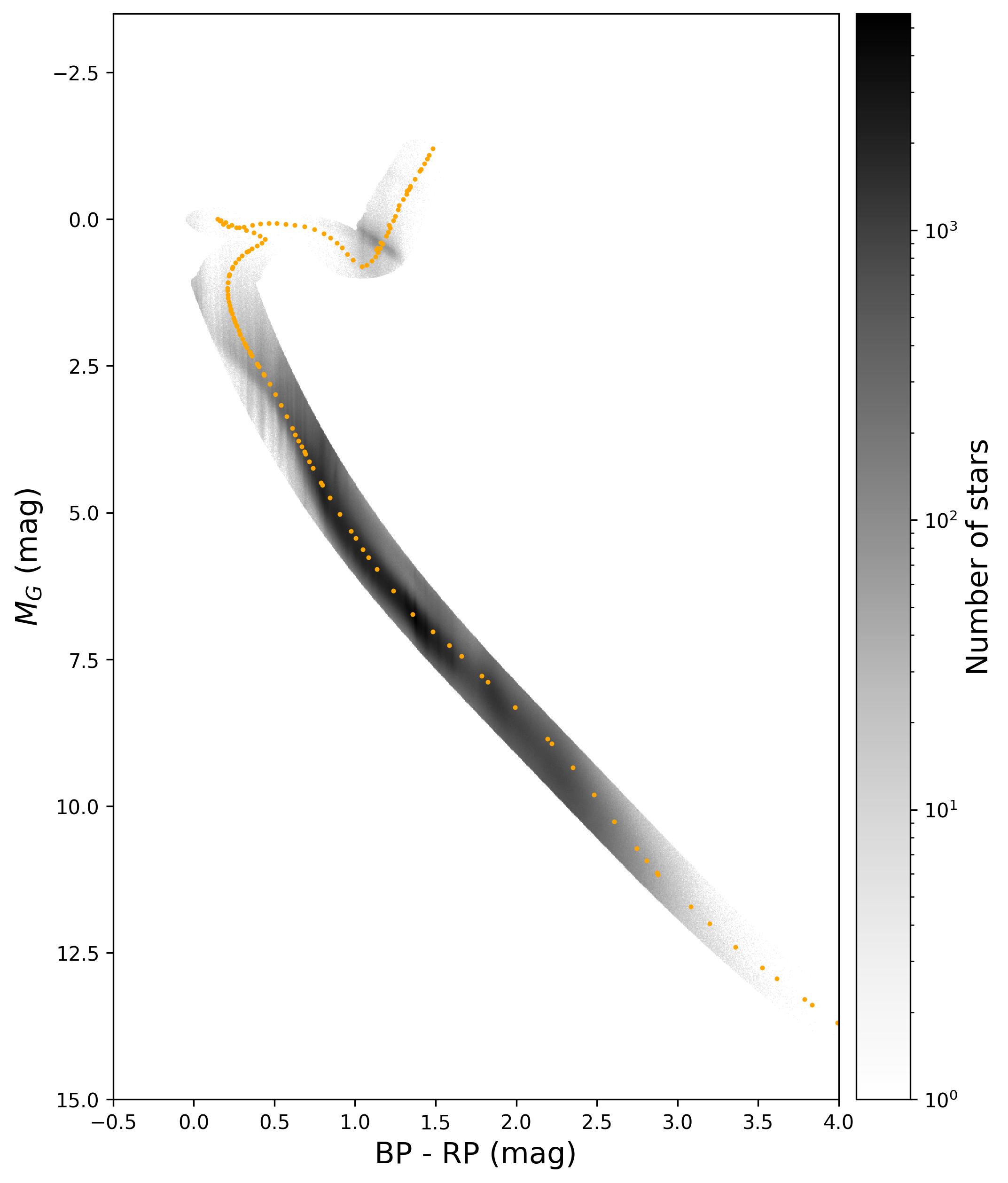}
    \caption{Gaia sources remaining after the isochrone filtering. The orange points represent the theoretical isochrone used for the selection.}
    \label{fig:isochrone}
\end{figure}

Subsequently, we refined our sample using an isochrone filter to mitigate contamination from unrelated sources, such as field dwarfs. We generated a PARSEC isochrone \cite{PARSEC..2012MNRAS.427..127B} with an age of 648\,Myr and a metallicity of $Z=0.02$. Our final selection retains only the stars located within a 0.2\,mag wide envelope around this isochrone. As illustrated in Fig.~\ref{fig:isochrone}, this step yields approximately 28 million stars. 

For stars that satisfied the isochrone selection criteria, we applied further spatial constraints to refine the sample. Specifically, we required the stars to be within $|x - x_{\mathrm{Hy}}| < 200\,\mathrm{pc}$, $|y - y_{\mathrm{Hy}}| < 500\,\mathrm{pc}$, and $|z - z_{\mathrm{Hy}}| < 20\,\mathrm{pc}$, where $(x_{\mathrm{Hy}}, y_{\mathrm{Hy}}, z_{\mathrm{Hy}})$ represents the central position of the Hyades cluster. This spatial filtering resulted in a sample of approximately $370\,000$ stars.

Subsequently, we constructed the corresponding CCP diagrams for the selected sample (hereafter referred to as ``Gaia CCP diagrams''). Note that because the linear parameters of the CCP method are model-dependent, a unique Gaia CCP diagram is generated for each specific spatially matched model. As shown in Fig.~\ref{fig:CCP}, the CCP diagrams of different models exhibit distinct substructures, which we initially intended to use as a diagnostic to further refine our model selection. Unfortunately, due to the substantial field star contamination, the Gaia CCP diagrams for all models only reveal a central clumpy structure, and the predicted internal substructures remain obscured. However, Fig.~\ref{fig:CCP} also indicates that, regardless of these substructures, the cluster members in each model generally reside within the kinematic windows defined by $|(V_{||,\mathrm{sim}} - V_{||,\mathrm{pred}}) - \Gamma(R - R_{\mathrm{cl}})| < 2\,\mathrm{km\,s^{-1}}$ and $|V_{||,\mathrm{pred}}| < 1.5\,\mathrm{km\,s^{-1}}$. Applying these selection cuts on Gaia CCP diagrams results in a refined sample of approximately 1,500 Gaia candidates for each spatial matched model.

\begin{figure*}[ht!]
    \centering
    \includegraphics[width=0.99\linewidth]{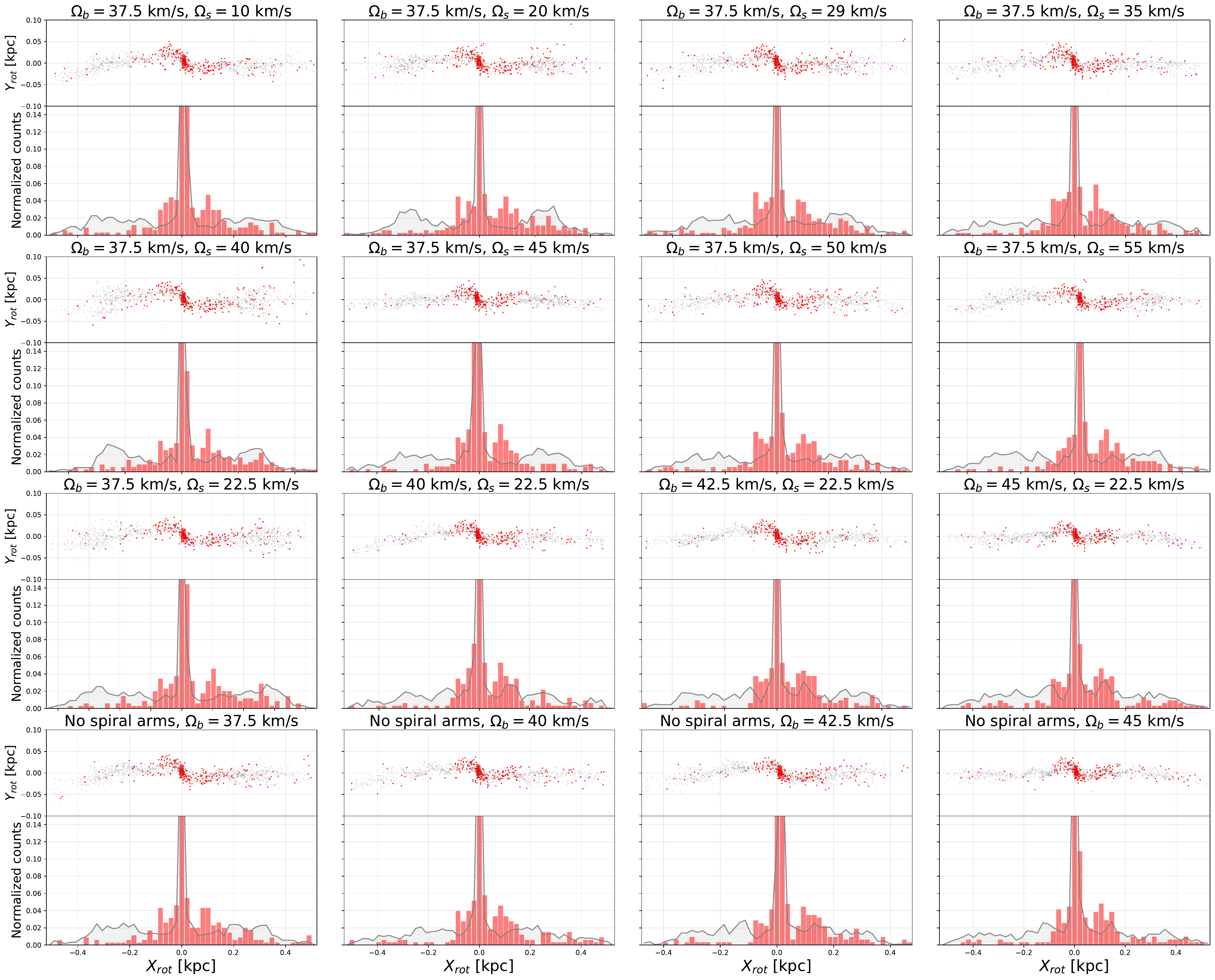}
    \caption{In each subplot, 
\textit{Upper panel:} Spatial distribution of the model (grey points) and the corresponding selected final members (red points). Both the model and final members are shown in the stream-aligned coordinate system $(X_{\mathrm{rot}}, Y_{\mathrm{rot}})$, where the $x_{\mathrm{rot}}$-axis is aligned with the principal direction of the stream's elongation. 
\textit{Lower panel:} Comparison of the longitudinal density distribution along the stream for the model (grey lines) and the final members (red histogram).}
    \label{fig:density_map}
\end{figure*}

Finally, we performed a spatial matching between the Gaia candidates and the model-predicted distributions. To achieve this, we employed a $k$-dimensional tree algorithm (\texttt{cKDTree} in \texttt{scipy}) to conduct a nearest-neighbor search in the three-dimensional Cartesian space $(x, y, z)$. To ensure the selection was robust against varying densities in the simulated streams, we adopted an adaptive distance threshold. Specifically, the threshold was set to $1.5$ times the 90th percentile of the internal nearest-neighbor distances among the model points. The resulting final members of each spatially matched model are shown as red dots in Fig.~\ref{fig:density_map}.

We further compared the longitudinal density distributions along the stream direction for the final members of each spatially matched model against their respective simulated density profiles. As illustrated in Fig.~\ref{fig:density_map}, the density distributions vary significantly across the different potential models. The locations of predicted over-densities differ between models, and in some cases, these features are even entirely absent. Theoretically, tidal streams originating from clusters on circular orbits within a static, axisymmetric potential are expected to exhibit periodic density enhancements, often referred to as epicyclic over-densities \citep{overdensity..2010MNRAS.401..105K, overdensity..2008MNRAS.387.1248K}. 
The suppression or "smearing out" of these structures in certain models highlights the impact of non-axisymmetric and time-dependent components, such as the Galactic bar and spiral arms, which perturb the stream's phase-space structure.

Notably, we find that nearly all the final members associated with different spatially matched models consistently exhibit a prominent density peak near $Y_{\mathrm{rot}} \approx 0.1\,\mathrm{kpc}$. Assuming this observed over-density accurately reflects the intrinsic physical structure of the Hyades stream, it suggests that potential models capable of reproducing this feature may provide a more realistic representation of the Galactic potential. This morphological signature is best recovered by models with $\Omega_s = 22.5\,\mathrm{km\,s^{-1}\,kpc^{-1}}$ and $\Omega_b = 40, 42.5$, and $45\,\mathrm{km\,s^{-1}\,kpc^{-1}}$. Significantly, these $\Omega_b$ are highly consistent with recent observational constraints that favor a relatively slow Galactic bar \citep{bar_speedP17..2017MNRAS.465.1621P}.

\section{Discussion and Limitations}\label{discussion}

Our comparison of the simulated and observed stream morphology and density structure favors models with $\Omega_b \simeq 40$--$45\,\mathrm{km\,s^{-1}\,kpc^{-1}}$ and $\Omega_s = 22.5\,\mathrm{km\,s^{-1}\,kpc^{-1}}$; here we discuss limitations and potential systematics affecting this inference.

A striking feature of the observed \textit{Gaia} data is the pronounced asymmetry between the two tidal tails, with one side appearing significantly more populated than the other. In contrast, while our simulated models under the influence of the bar and spiral arms do exhibit some degree of asymmetry, they do not fully reproduce the near-disappearance of one tail seen in the observations. A possible explanation is that the high level of field star contamination near the Galactic disk makes it extremely difficult to identify members in the more diffuse regions of the stream. The lack of radial velocity (RV) data for the majority of the candidate stars in the distant tails makes it difficult to uniquely distinguish true stream members from the background population. Future large-scale spectroscopic surveys like WEAVE \citep{weave..2012SPIE.8446E..0PD} and 4MOST \citep{4most..2019Msngr.175....3D} will be crucial for confirming the chemical and kinematic signatures of the distant tidal tail members. Integrating these 6D phase-space data with chemical tagging will allow for a more robust test of the Galactic potential models presented in this work. Alternatively, the observed asymmetry may have a physical origin. A significant past collision or a series of close encounters with GMCs could have led to the structural disruption or erasure of one side of the stream \citep{stream_with_collision..2006MNRAS.371..793G, J21..2021A&A...647A.137J}.

Beyond these observational constraints and potential stochastic events, the results of this study are also subject to several inherent simplifications and uncertainties in our numerical modeling approach. In this study, we varied $\Omega_b$ while keeping its length and strength constant. In reality, the bar's length is often correlated with $\Omega_b$, with slower bars typically being longer \citep{bar_length..2003MNRAS.341.1179A}. Furthermore, the Galactic bar and spiral arms may undergo secular evolution, meaning their pattern speeds could be time-dependent \citep{changing_bar_speed..2020MNRAS.497..933H}. By adopting a static bar geometry and $\Omega_b$, our model might slightly shift the exact resonance positions, which could influence the predicted stream morphology.

In addition, the intrinsic parameters of the Hyades cluster, specifically its age and initial mass, also introduce uncertainties. While we adopted an age of $648\,\mathrm{Myr}$ and an initial mass of $800\,M_{\odot}$ based on recent literature \citep{WangMW2014..2021A&A...655A..71W}, different age estimates ranging from $600$ to $800\,\mathrm{Myr}$ \citep[e.g.,][]{early_Hyades..1998A&A...331...81P, Hyades_age..2009ApJ...696...12D, Hyades_age..2018MNRAS.477.3197R, Gaia..2018A&A...616A..10G} would directly affect the simulated length of the tidal tails. Similarly, the exact solar position ($R_0$) and velocity ($v_{\odot}$) are still subjects of refinement. Since the transformation from ICRS to Galactocentric coordinates is highly sensitive to these solar parameters, small variations in $(R_0, V_{\odot})$ could lead to systematic shifts in the estimated orbit of the Hyades \citep{Sun_position..2019ApJ...885..131R, SUn_distance..2022A&A...657L..12G}.

Furthermore, determining the precise initial phase-space coordinates of the cluster $648\,\mathrm{Myr}$ ago remains a challenge. Although we utilized the Nelder-Mead optimization method to refine the backward-integrated orbit, the chaotic nature of orbits in a non-axisymmetric potential means that the ``initial guess'' may not be unique. The resulting trajectory is a best-fit solution, but the sensitivity of the tidal tail formation to the exact orbital path suggests that these initial conditions remain a significant source of uncertainty. 

\section{Conclusions}

We modeled the formation of the Hyades tidal stream with direct $N$-body simulations (\texttt{PETAR}) in a multi-component Milky Way potential implemented with \texttt{AGAMA} and including a rotating bar and spiral arms. By systematically varying the bar and spiral pattern speeds, we quantified how non-axisymmetric forcing reshapes the stream orientation, length, and internal density structure over the $\sim 648\,\mathrm{Myr}$ evolution of the cluster. We then confronted the models with observations by combining CP/CCP-based kinematic selection on \textit{Gaia} EDR3 with an adaptive three-dimensional spatial matching technique.

The key observational constraint is the stream-aligned longitudinal density profile of the \textit{Gaia} candidates. This profile shows a prominent peak at $Y_{\mathrm{rot}} \approx 0.1\,\mathrm{kpc}$, which is consistent with the overdensity reported by \citet{J21..2021A&A...647A.137J}. We find that this feature is best reproduced by models with $\Omega_s = 22.5\,\mathrm{km\,s^{-1}\,kpc^{-1}}$ and $\Omega_b \simeq 40$--$45\,\mathrm{km\,s^{-1}\,kpc^{-1}}$, supporting a relatively slow Galactic bar. These findings also indicate that the inclusion of spiral arms is not negligible and $\Omega_s$ around this intermediate value are favored. More broadly, our results demonstrate that nearby open-cluster streams provide a complementary, local avenue to constrain non-axisymmetric Galactic dynamics using not only the global stream orientation but also its fine-grained density structure.

\begin{acknowledgments}
LW thanks the support from the National Natural Science Foundation of China through grant 12573041 and 12233013, the High-level Youth Talent Project (Provincial Financial Allocation) through the grant 2023HYSPT0706, the Fundamental Research Funds for the Central Universities, Sun Yat-sen University (2025QNPY04). TJ acknowledges the support from the MUNI Award in Science and Humanities (MUNI/I/1762/2023).
\end{acknowledgments}

\begin{contribution}
ZZ was responsible for the method development, performing the simulations, data analysis, and writing the manuscript.
LW camp up with the initial research concept, contributed to discussions, edited the manuscript, and supervised the overall project.
TJ developed the observational-comparison concept, provided observational data, contributed to discussions, and edited the manuscript.
ZH contributed to the Galactic-potential method development and discussions.

\end{contribution}

\software{\texttt{NumPy} (\citealp{85-Harris.2020}),
          \texttt{Matplotlib} (\citealp{86-Hunter.2007}),
          \texttt{Astropy} (\citealp{astropy2013..2013A&A...558A..33A, astropy2018..2018AJ....156..123A, astropy2022..2022ApJ...935..167A}),
          \texttt{SDAR} (\citealp{Petar_SDAR..2020MNRAS.491.2413W}, https://github.com/lwang-astro/SDAR),
          \texttt{PETAR} (\citealp{Petar..2020MNRAS.497..536W}, https://github.com/lwang-astro/PeTar),
          \texttt{FDPS} (\citealp{69-Iwasawa.2016,70-Iwasawa.2020,71-Namekata.2018}, https://github.com/FDPS/FDPS),
          \texttt{SSE/BSE} (\citealp{Hurley2000,Hurley2002,Banerjee2020}),
          \texttt{McLuster} (\citealp{mcluster..2011MNRAS.417.2300K,Wang2019}, modified version: https://github.com/lwang-astro/mcluster),
          \texttt{AGAMA} (\citealp{agama..2019MNRAS.482.1525V}, https://github.com/GalacticDynamics-Oxford/Agama), 
          \texttt{Astroquery} (\citealp{astroquery..2019AJ....157...98G})}
          
\bibliography{sample701}{}
\bibliographystyle{aasjournalv7}
\end{document}